
\input epsf.tex

\font\rmu=cmr10 scaled\magstephalf
\font\bfu=cmbx10 scaled\magstephalf

\font\it=cmti10 scaled \magstephalf
\font\bf=cmbx10 scaled\magstephalf
\rmu

\font\rmus=cmr8
\font\rmuss=cmr6
\font\mait=cmmi10 scaled\magstephalf
\font\maits=cmmi7 scaled\magstephalf
\font\maitss=cmmi7
\font\msyb=cmsy10 scaled\magstephalf
\font\msybs=cmsy8 scaled\magstephalf
\font\msybss=cmsy7
\font\bfus=cmbx7 scaled\magstephalf
\font\bfuss=cmbx7
\font\cmeq=cmex10 scaled\magstephalf

\textfont0=\rmu
\scriptfont0=\rmus
\scriptscriptfont0=\rmuss

\textfont1=\mait
\scriptfont1=\maits
\scriptscriptfont1=\maitss

\textfont2=\msyb
\scriptfont2=\msybs
\scriptscriptfont2=\msybss

\textfont3=\cmeq
\scriptfont3=\cmeq
\scriptscriptfont3=\cmeq

\newfam\bmufam  \textfont\bmufam=\bfu
      \scriptfont\bmufam=\bfus \scriptscriptfont\bmufam=\bfuss

\hsize=15.5cm
\vsize=21cm
\baselineskip=16pt   
\parskip=12pt plus  2pt minus 2pt

\def\semi{\bigcirc\kern-1em{s}\;}

\def\R{{\rm I\!R}}

\def\one{{\mathchoice {\rm 1\mskip-4mu l} {\rm 1\mskip-4mu l}
{\rm 1\mskip-4.5mu l} {\rm 1\mskip-5mu l}}}
\def\Q{{\mathchoice
{\setbox0=\hbox{$\displaystyle\rm Q$}\hbox{\raise 0.15\ht0\hbox to0pt
{\kern0.4\wd0\vrule height0.8\ht0\hss}\box0}}
{\setbox0=\hbox{$\textstyle\rm Q$}\hbox{\raise 0.15\ht0\hbox to0pt
{\kern0.4\wd0\vrule height0.8\ht0\hss}\box0}}
{\setbox0=\hbox{$\scriptstyle\rm Q$}\hbox{\raise 0.15\ht0\hbox to0pt
{\kern0.4\wd0\vrule height0.7\ht0\hss}\box0}}
{\setbox0=\hbox{$\scriptscriptstyle\rm Q$}\hbox{\raise 0.15\ht0\hbox to0pt
{\kern0.4\wd0\vrule height0.7\ht0\hss}\box0}}}}
\def\C{{\mathchoice
{\setbox0=\hbox{$\displaystyle\rm C$}\hbox{\hbox to0pt
{\kern0.4\wd0\vrule height0.9\ht0\hss}\box0}}
{\setbox0=\hbox{$\textstyle\rm C$}\hbox{\hbox to0pt
{\kern0.4\wd0\vrule height0.9\ht0\hss}\box0}}
{\setbox0=\hbox{$\scriptstyle\rm C$}\hbox{\hbox to0pt
{\kern0.4\wd0\vrule height0.9\ht0\hss}\box0}}
{\setbox0=\hbox{$\scriptscriptstyle\rm C$}\hbox{\hbox to0pt
{\kern0.4\wd0\vrule height0.9\ht0\hss}\box0}}}}

\font\fivesans=cmss10 at 4.61pt
\font\sevensans=cmss10 at 6.81pt
\font\tensans=cmss10
\newfam\sansfam
\textfont\sansfam=\tensans\scriptfont\sansfam=\sevensans\scriptscriptfont
\sansfam=\fivesans
\def\sans{\fam\sansfam\tensans}
\def\Z{{\mathchoice
{\hbox{$\sans\textstyle Z\kern-0.4em Z$}}
{\hbox{$\sans\textstyle Z\kern-0.4em Z$}}
{\hbox{$\sans\scriptstyle Z\kern-0.3em Z$}}
{\hbox{$\sans\scriptscriptstyle Z\kern-0.2em Z$}}}}

\newcount\foot
\foot=1
\def\note#1{\footnote{${}^{\number\foot}$}{\ftn #1}\advance\foot by 1}

\def\frac#1#2{{#1\over #2}}
\def\text#1{\quad{\hbox{#1}}\quad}

\font\ch=cmbx12 scaled\magstephalf
\font\ftn=cmr8 scaled\magstephalf

\font\it=cmti10 scaled\magstephalf
\font\bf=cmbx10 scaled\magstephalf
\font\titch=cmbx12 scaled\magstep2
\font\titname=cmr10 scaled\magstep2
\font\titit=cmti10 scaled\magstep1
\font\titbf=cmbx10 scaled\magstep2

\nopagenumbers

\line{\hfil AEI-028}
\line{\hfil February 26, 1997}
\vskip2cm
\centerline{\titch Imposing det E $>$ 0 in discrete}
\vskip.5cm
\centerline{\titch quantum gravity}
\vskip1.2cm
\centerline{{\titname R. Loll}\footnote*{e-mail: loll@aei-potsdam.mpg.de}}
\vskip1cm
\centerline{\titit Max-Planck-Institut f\"ur Gravitationsphysik}
\vskip.2cm
\centerline{\titit Schlaatzweg 1}
\vskip.2cm
\centerline{\titit D-14473 Potsdam, Germany}
\vskip.3cm

\vskip2.5cm
\centerline{\titbf Abstract}
\vskip0.2cm
We point out that the inequality $\det E>0$ distinguishes
the kinematical phase space of canonical connection gravity from that
of a gauge field theory, and characterize the eigenvectors with
positive, negative and zero-eigenvalue of the corresponding quantum
operator in a lattice-discretized version of the theory. 
The diagonalization of $\hat {\det E}$ is 
simplified by classifying its eigenvectors according to the 
irreducible representations of the octagonal group.

\noindent keywords: canonical quantum gravity, determinant of the metric,
lattice gravity, volume operator, octagonal group
\vfill\eject
\footline={\hss\tenrm\folio\hss}
\pageno=1

\line{\ch 1 Introduction\hfil}

It is sometimes stated that the unconstrained phase space of pure gravity
in the Ashtekar formulation [1] is that of a Yang-Mills theory. This 
however is not quite true. The origin of this subtlety has nothing to do
with the complexification of the connection form in the original formulation,
and is indeed also present in the purely real connection formulation [2],
which is the subject of this letter. 

Recall that in this Hamiltonian form of Lorentzian gravity, the basic
canonical variable pair $(A_a^i,E^a_i)$ consists of an $su(2)$-valued 
gauge potential $A$ and a densitized, inverse dreibein $E$. Denoting
the dreibein (the ``square root of the three-metric") by $e_a^i$, 
$e_a^ie_{bi}=g_{ab}$, with its inverse $e^a_i$ satisfying $e^a_i e_a^j=
\delta_i^j$, $E$ can be expressed as $E^a_i=(\det e_b^j)e^a_i$.
Taking the determinant of this equation, one obtains
$\det E^a_i=(\det e_b^j)^2$, which (for non-degenerate metrics) is always
positive and non-vanishing ($\det e_b^j$ alone may assume
values $\pm \sqrt{\det g_{ab}}$). 

However, once one chooses the $E^a_i$'s as the basic variables, the
inequality $\det E>0$ {\it has to be imposed as an extra condition} 
to recover the
correct gravitational phase space. This is analogous to Hamiltonian
metric formulations for gravity where the condition $\det g>0$ must
be imposed on the symmetric 3-tensors $g_{ab}$, constituting half
of the canonical variables. Similar conditions also appear in other
gauge-theoretic reformulations of gravity. 
One crucial question is how such a condition is to be translated
to the quantum theory. Fortunately this is possible in the case of
connection gravity, at least in a lattice-discretized version of the
theory. 
  
If one quantizes connection gravity along the lines of a non-abelian 
gauge field theory, as is usually done, and
as is suggested by the kinematical resemblance of the two,
an operator condition like $\hat {\det E}>0$ is {\it not} automatically 
satisfied. Since $\det E$ is classically a third-order polynomial 
in the momenta $E^{a}_{i}$,

$$
\det E =\frac{1}{3!} \eta_{abc}\epsilon^{ijk}E_{i}^{a}E_{j}^{b}E_{k}^{c},
\eqno(1.1)
$$

\noindent and since in the Yang-Mills-like quantization the momenta are 
represented by $i$ times diffe\-ren\-tiation with respect to $A$, 
$\hat {\det E}\,\Psi>0$ 
is a differential condition for physical wave functions $\Psi$, and an
obvious candidate for a quantization of the classical inequality 
$\det E>0$.  
There already exists a well-defined, self-adjoint lattice operator
with discrete spectrum, 
which is the quantized version of a discretization of the classical
function $\det E$ [3]. We call this 
operator $\hat D(n)$, where $n$ labels the vertices of a 
three-dimensional lattice with cubic topology, and $\hat D(n)$ is 
written in terms of the symmetrized link momenta $\hat p(n,\hat a)$
as

$$
\hat D(n):=\frac{1}{3!} \eta_{\hat a \hat b \hat c}\epsilon^{ijk}
\hat p_{i}(n,\hat a) \hat p_{j}(n,\hat b) \hat p_{k}(n,\hat c),
\eqno(1.2)
$$
\noindent where
$$
\hat p_i(n,\hat a)=\frac{i}{2}(X_+^i(n,\hat a)+X_-^i(n-1_{\hat a},\hat a)),
\eqno(1.3)
$$

\noindent and $X_\pm^i(n,\hat a)$ denote the left- and right-invariant
vector fields on the group manifold associated with the link $l=(n,\hat a)$,
with commutators $[X_{\pm}^{i},X_{\pm}^{j}]=\pm\epsilon^{ijk}X_{\pm}^{k}$.
(For convenience we have rescaled $D(n)$ by a factor of $\frac{1}{6}$
with respect to the definition in [3].)
The square root of $\hat D(n)$ (whenever it is defined) 
is the so-called volume
operator, and some of its spectral properties have been investigated
both on the lattice and in the continuum. The latter is relevant
because it turns out that self-adjoint volume operators can be defined in the
continuum loop representation of quantum gravity [4,5,6]. After 
regularization their action on fixed, imbedded spin network states is
very much like that of a lattice operator. In parti\-cu\-lar, the finite
volume operators of [5,6] (up to overall factors
and modulus signs) coincide on suitable geometries
with (1.2) (this is explained in more detail in [7]). 
The volume operator and its discretized version have emerged as 
important ingredients in the construction of the quantum Hamiltonian
constraint. Note that the non-polynomial quantities
appearing in canonical connection gravity can always be rewritten
in polynomial form modulo arbitrary powers of $\det E$. Thus, if one
can explicitly quantize $\det E$, arbitrary functions of $\det E$
can be quantized in terms of its spectral resolution. If inverse powers
of $\det E$ appear, one in addition has to identify the
zero-eigenstates of $\hat {\det E}$ [8]. 

There is therefore clearly a need for a better understanding of the spectral
properties of the operator $\hat {\det E}$.
There exist general formulae for its matrix elements, obtained in various
preferred orthogonal bases of wave functions [4,6].
Since one does not expect to be able to establish general analytic 
formulae for the spectrum itself, the limits for evaluating
it numerically are given by the size of the matrices that are to be
diagonalized and the computing power available. We will below describe
a way of reducing the matrix size, by establishing a set of superselection
sectors on which $\hat D(n)$ can be diagonalized separately. They have
their origin in discrete geometric symmetries of the operator and the Hilbert 
space on which it is defined.

Our discussion will take place within the lattice theory, but for the
reasons mentioned above, results about the lattice spectrum translate,
at least partially, into results about the continuum spectrum.

\vskip1.5cm
\line{\ch 2 Characterization of eigenstates\hfil}

It was already noted during earlier investigations of the volume
spectrum [9,3] that non-vanishing eigenvalues of the operator
$\hat{\det E}$ always appear in pairs of opposite sign. That this is
also true in general can be seen as follows (the argument is similar to
the one used to prove that three-valent spin network states necessarily
have vanishing volume [9]). 
We work on the gauge-invariant sector ${\cal H}^{\rm inv}$ 
of the lattice gauge-theoretic
Hilbert space, whose elements are linear combinations of Wilson loops,
i.e. of traces of closed lattice holonomies.
A convenient way of labelling a basis of states is given by $|j_l,\vec v_n>$
(so-called spin network states),
where $j_l=0,1,2,\dots$ labels the $su(2)$-representation associated with 
each lattice link $l=(n,\hat a)$, and $\vec v_n$ is a set of linearly
independent intertwiners (contractors of Wilson lines) compatible with
the $j_l$ at each lattice vertex $n$. Note that these states are real
functions of the $SU(2)$-lattice holonomies.

Since $\hat D(n)$ only acts locally at $n$, we need only consider 
the part of Hilbert space associated 
with the single vertex $n$ and the six links intersecting at $n$.
Moreover, $\hat D(n)$ leaves the flux line numbers $j_l$ alone,
and therefore acts non-trivially only on the finite-dimensional
spaces of the linearly independent intertwiners labelled by $\vec v_n$.

Consider an orthogonal basis of states $\{\phi_i\}$ in one of these
finite-dimensional spaces, 
and assume that $\Psi$ is an eigenstate of $\hat D(n)$,
$\hat D(n)\Psi =d\Psi$. Since $\hat D(n)$ is a self-adjoint operator,
$d$ is a real number. 
In this basis, the decomposition for $\Psi$ reads
$\Psi=\sum_i (a_i+i b_i)\phi_i$, $a_i,b_i,\in\R$. Since the explicit operator
expression for $\hat D(n)$ is purely imaginary, as can be seen from (1.2,3),
it immediately follows from

$$
\hat D(n)\sum_i (a_i+i b_i)\phi_i=d\ \sum_i (a_i+i b_i)\phi_i,
\eqno(2.1)
$$

\noindent by taking the complex conjugate that

$$
\hat D(n)\sum_i (a_i-i b_i)\phi_i=-d\ \sum_i (a_i-i b_i)\phi_i.
\eqno(2.2)
$$

\noindent The consequences can be summarized as follows: if $\Psi$ is
an eigenstate of $\hat D(n)$ with eigenvalue $d$, then its complex conjugate
$\Psi^*$ is also an eigenstate, with eigenvalue $-d$. If an eigenstate
$\Psi$ is a purely real or a purely imaginary linear combination of
spin network states, then its eigenvalue must necessarily be $d=0$.

This provides a first characterization of positive-, negative- and
zero-eigenstates of the operator $\hat D(n)$. 
That there should be such a one-to-one map between
states of positive and negative volume is plausible from a physical
point of view, since for Yang-Mills configurations there is no
preferred orientation for triples of $E$-fields. 

A first practical consequence for the computation of spectrum and
eigenstates of $\hat D$ is the following. Although $\hat D$ does not
commute with complex conjugation, its square $\hat D^2$ (which also
is a well-defined self-adjoint operator) does. Therefore $\hat D^2$ can
be diagonalized already on the subspace of real states.
Assume now that $\chi$ is such a real eigenstate of $\hat D^2$,
$\hat D^2(n)\chi=v^2\chi$, $v\not=0$. It follows immediately that its image
under $\hat D$ is an (imaginary) eigenstate of $\hat D^2$ since
$\hat D^2(\hat D\chi)=v^2 \hat D\chi$. 
Consider the linear combination of these two states under the action
of $\hat D$,
 
$$
\hat D(\chi\pm\frac{1}{|v|}\hat D \chi)=\pm\frac{1}{|v|}\hat D^2\chi+
\hat D\chi=\pm |v| (\chi \pm \frac{1}{|v|}\hat D\chi).
\eqno(2.3)
$$

\noindent Thus, we can read off a recipe for constructing positive volume
eigenstates: take any eigenstate $\chi$ of $\hat D^2$ with non-zero eigenvalue
$v^2$, then $\chi +\frac{1}{|v|}\hat D \chi$ is an eigenvector of $\hat D$
with eigenvalue $|v|$.

\vskip1.5cm
\line{\ch 3 The role of the octagonal group\hfil}

In order to simplify the task of finding eigenstates of $\hat D$, 
we will construct operators that commute with it and among themselves,
and can therefore be diagonalized simultaneously. The finite-dimensional
matrices associated with the action of $\hat D$ on
vertex states of given flux line numbers 
decompose into block-diagonal form, and the blocks can be
diagonalized individually. 

The key observation is that the classical lattice function $D(n)\equiv
{\det E}(n)$ is invariant under the action of the discrete group $\cal O$
of 24 elements, called the octagonal or cubic group [10]. They can be thought
of as the permutations of the three (oriented) lattice axes meeting at the
intersection $n$ which do not change the orientation of the local coordinate
system they define. By contrast, $D(n)$ changes sign under the total
space reflection $T$ (i.e. under simultaneous inversion of the three axes).  
It is sometimes convenient to consider the discrete group of 48 elements
${\cal O}\times T$.

As a result of this classical symmetry, 
eigenstates of $\hat D(n)$ can be classified according
to the irreducible representations of $\cal O$. 
This set-up is familiar to lattice gauge theorists,
because it has been employed in analyzing
the glueball spectrum of the Hamiltonian in four-dimensional
$SU(3)$-lattice gauge theory [11]. Adapted to the present $SU(2)$-context,
certain further simplifications occur which have to do with how
the gauge-invariant sector of the lattice theory is labelled by the
spin network states. 

One way of labelling local  
spin network states at a vertex $n$ is the following.
Fix a local coordinate system at $n$ and label the three incoming links
as $(-1,-2,-3)$, and the corresponding outgoing ones as $(1,2,3)$,
and the corresponding link fluxes by $j_i$, $i=\pm 1,2,3$.
(The $j_i$ cannot be chosen totally freely but must be such that suitable
gauge-invariant routings of flux lines through the intersection exist.)
To take care of the intertwiners,
call $j_{m,n}$ the number of spin-$\frac{1}{2}$-flux lines coming in
at link $m$ and going out at link $n$. Both $m$ and $n$ can take positive
and negative values, but $m= n$ is excluded, since it corresponds
to a trivial retracing of a link. Since the flux lines appearing
in spin network states are not sensitive to orientation, there
are 15 numbers $j_{m,n}$. They are subject to a number of
constraints since the total number of flux lines $j_i$ associated with a
given incoming or outgoing link is assumed fixed. 
Our reason for choosing this label set for the contractors is
their simple transformation behaviour under the cubic group.

This way of labelling still contains a large
redundancy in the form of so-called Mandelstam constraints. This is
partially eliminated by choosing a smaller label set: again fix an orientation
of the three axes, and consider only intertwiners with non-vanishing
$\{j_{-1,1},j_{-1,2},j_{-1,3},$ $j_{-2,1},j_{-2,2},j_{-2,3},
j_{-3,1},j_{-3,2},j_{-3,3}\}$. It can easily be shown that all other 
intertwiners can be written
as linear combinations of this set, by virtue of the Mandelstam
identities. Moreover, $\hat D(n)$ maps the set into itself. However,
the symmetry group $\cal O$ does not leave it invariant;
only a six-dimensional subgroup (which we will call ${\cal O}^{(6)}$)
maps the set into itself. Dropping the minus signs in front of the
negative subscripts of the $j_{mn}$ in the reduced 9-element set, 
let us rearrange the data in a $3\times 3$-matrix $J$,

$$
J:=
\left(\matrix{j_{11}&j_{12}&j_{13}\cr j_{21}&j_{22}&j_{23}\cr
j_{31}&j_{32}&j_{33}\cr}\right).
\eqno(3.1)
$$

The non-trivial elements of ${\cal O}^{(6)}$ in this notation
are represented by 

$$
\eqalign{
&R_1(J):=\left(\matrix{j_{11}&j_{31}&j_{21}\cr j_{13}&j_{33}&j_{23}\cr
  j_{12}&j_{32}&j_{22}\cr}\right),\;
R_2(J):=\left(\matrix{j_{33}&j_{23}&j_{13}\cr j_{32}&j_{22}&j_{12}\cr
  j_{31}&j_{21}&j_{11}\cr}\right),\;
R_3(J):=\left(\matrix{j_{22}&j_{12}&j_{32}\cr j_{21}&j_{11}&j_{31}\cr
  j_{23}&j_{13}&j_{33}\cr}\right),\cr
&\hskip2cm S_1(J):=\left(\matrix{j_{22}&j_{23}&j_{21}\cr 
  j_{32}&j_{33}&j_{31}\cr
  j_{12}&j_{13}&j_{11}\cr}\right),\;
S_2(J):=\left(\matrix{j_{33}&j_{31}&j_{32}\cr j_{13}&j_{11}&j_{12}\cr
  j_{23}&j_{21}&j_{22}\cr}\right).}
\eqno(3.2)
$$

\noindent We will also use the total space reflection $T$,

$$
T(J):=\left(\matrix{j_{11}&j_{21}&j_{31}\cr j_{12}&j_{22}&j_{32}\cr
  j_{13}&j_{23}&j_{33}\cr}\right).
\eqno(3.3)
$$

\noindent Since $T$ commutes with all elements of ${\cal O}^{(6)}$,
adjoining it we obtain a 12-element group 
${\cal O}^{(6)} \times T \equiv {\cal O}^{(6)} \times\Z_2$.
The multiplication table for the group ${\cal O}^{(6)}$ is
given in Table 1.

\vskip.7cm
{\offinterlineskip\tabskip=0pt
 \halign{ \strut\vrule#& \quad # \quad &\vrule#&
          \quad\hfil #\quad &
          \quad\hfil #\quad &
          \quad\hfil #\quad &
          \quad\hfil #\quad &
          \quad\hfil #\quad &
          \quad\hfil #\quad &\vrule#\cr \noalign{\hrule}
 & & & & & & && &\cr
 & && $\one $ & $R_1$ & $R_2$
              & $R_3$ & $S_1$ & $S_2$ &\cr
 & & & & & & && &\cr
 \noalign{\hrule}
 & & & & & & && &\cr
 & $\one$ &&
      $\one$ & $R_1$ & $R_2$ & $R_3$ & $S_1$ & $S_2$ &\cr
 & & & & & & && &\cr
 & $R_1$ &&
      $R_1$ & $\one$ & $S_1$ & $S_2$ & $R_2$ & $R_3$ &\cr
 & & & & & & && &\cr
 & $R_2$ &&
      $R_2$ & $S_2$ & $\one$ & $S_1$ & $R_3$ & $R_1$ &\cr
 & & & & & & && &\cr
 & $R_3$ &&
      $R_3$ & $S_1$ & $S_2$ & $\one$ & $R_1$ & $R_2$ &\cr
 & & & & & & && &\cr
 & $S_1$ &&
      $S_1$ & $R_3$ & $R_1$ & $R_2$ & $S_2$ & $\one$ &\cr
 & & & & & & && &\cr
 & $S_2$ &&
      $S_2$ & $R_2$ & $R_3$ & $R_1$ & $\one$ & $S_1$ &\cr
 & & & & & & && &\cr
 \noalign{\hrule} }
\normalbaselines
\baselineskip=16pt   
\vskip.5cm
\line{{\bf Table 1} \hskip.3cm Multiplication table for the subgroup 
${\cal O}^{(6)}$ of the octagonal group.\hfill}
\vskip.7cm

It is easy to generate all allowed intertwiner configurations $J$, given
flux line assignments $j_i$, $i=-1,-2,-3,1,2,3$, 
for the in- and outgoing links. The elements of the rows and columns
of $J$ simply have to add up to the appropriate $j_i$, for example,
$\sum_{i=1}^3 j_{1,i}=j_{-1}$, $\sum_{i=1}^3 j_{i,1}=j_{1}$.
Another advantage of this form is that the still remaining Mandelstam
constraints can be expressed as simple linear combinations of
$J$-matrices, and are all of the form

$$
\eqalign{
&\left(\matrix{j_{11}+1&j_{12}&j_{13}\cr j_{21}&j_{22}+1&j_{23}\cr
  j_{31}&j_{32}&j_{33}+1\cr}\right)
-\left(\matrix{j_{11}+1&j_{12}&j_{13}\cr j_{21}&j_{22}&j_{23}+1\cr
  j_{31}&j_{32}+1&j_{33}\cr}\right)\cr
-&\left(\matrix{j_{11}&j_{12}+1&j_{13}\cr j_{21}+1&j_{22}&j_{23}\cr
  j_{31}&j_{32}&j_{33}+1\cr}\right)+
\left(\matrix{j_{11}&j_{12}+1&j_{13}\cr j_{21}&j_{22}&j_{23}+1\cr
  j_{31}+1&j_{32}&j_{33}\cr}\right)\cr
+&\left(\matrix{j_{11}&j_{12}&j_{13}+1\cr j_{21}+1&j_{22}&j_{23}\cr
  j_{31}&j_{32}+1&j_{33}\cr}\right)
-\left(\matrix{j_{11}&j_{12}&j_{13}+1\cr j_{21}&j_{22}+1&j_{23}\cr
  j_{31}+1&j_{32}&j_{33}\cr}\right)=0}\eqno(3.4)
$$
 
\noindent Obviously, (3.4) is not to be understood as a matrix equation; 
the matrices are only labels for Hilbert space elements.
The operator $\hat D$ is cubic in derivatives, and can therefore
be written as a sum of terms, each of which acts on some triplet
of spin-$\frac{1}{2}$ flux lines routed through the intersection $n$.
Its explicit form can be derived in a straightforward way, but
is too long to be reproduced
here. It can be found in our forthcoming publication [12]. Its form
is a linear combination (with $j_{mn}$-dependent coefficients)
of matrices $J$ whose entries differ at
most by $\Delta j_{mn}=\pm 1$ from the input matrix. This gives us a
general formula for matrix elements, albeit in a non-orthogonal basis.

One finds the following relations under conjugation with elements of
${\cal O}^{(6)} \times T$:

$$
R_i\hat D R_i=\hat D,\; i=1,2,3,\;\;
S_i\hat D S_i=\ \hat D,\; i=1,2,\;\;
T\hat D T=-\hat D.
\eqno(3.5)
$$

Next, we are interested in the representation theory of these discrete
groups. ${\cal O}^{(6)}$ contains three conjugacy classes of elements
namely, $\{\one\},\{R_1,R_2,R_3\}$ and $\{S_1,S_2\}$. Following [10],
one establishes the existence of three irreducible representations: two 
one-dimensional ones (called $A_1$ and $A_2$) and one two-dimensional one
(called $E$). They can be identified by the values of their characters,
i.e. the traces of the matrices representing the group elements (which
only depend on the conjugacy class). The enlarged group
${\cal O}^{(6)}\times T$ has six conjugacy classes and six irreducible
representations, since each of the previous representations gives rise
to one of positive and one of negative parity, denoted by a subscript
$+$ or $-$.
The possible orbits sizes through single elements $J$ under the
action of ${\cal O}^{(6)}\times T$ are 1, 2, 3, 6 and 12, and they
have a well-defined irreducible representation content [12].

It follows from (3.5) that $\hat D$ obeys the (anti-)commutation relations

$$
[\hat D,R_i]=0,\;i=1,2,3,\;\; 
[\hat D,S_1+S_2]=0,\;
[\hat D,T]_+=0. 
\eqno(3.6)
$$

We conclude that $\hat D$ does not alter the ${\cal O}^{(6)}$-quantum
numbers, but maps positive- into negative-parity states and vice versa. 
In practice it is convenient to work with the operator $\hat D^2$. 
A maximal subset of operators commuting both among themselves and with
$\hat D^2$ is, for example, $\{R_1+R_2+R_3,S_1+S_2,T\}$.
This of course implies that $\hat D^2$ may be diagonalized separately on
the eigenspaces of these operators, reducing the problem to a
smaller one. 

One further observation turns out to be useful. Since
parity-odd wave functions are constructed by weighted sums (with
factors $\pm1$) of spin network states, which may sometimes vanish,
there are always fewer states transforming according to the representations
$A_i^-$, $E^-$, than those transforming according to $A_i^+$, $E^+$.
The most efficient way of diagonalizing $\hat D$ is therefore to start
from the set of wave functions transforming according to one of the
negative-parity irreducible representations, diagonalize $\hat D^2$, 
construct the images under $\hat D$ of the resulting set of states
(which all have positive parity), and then form complex linear
combinations to obtain eigenstates of $\hat D$, as explained in the
previous section. The number of zero-volume states is then given
by the difference of positive- and negative-parity states.

As an application of this scheme, we have analyzed the irreducible
representation content of some of the Hilbert spaces corresponding to
flux line numbers $(j_{-1},j_{-2},j_{-3},j_1,j_2,j_3)=
(j,j,j,j,j,j)$, i.e. for genuine six-valent intersections [12]. 
In this case, ${\cal O}^{(6)}\times T$ maps the
Hilbert space into itself. Matrix sizes are reduced considerably
when the various super\-selection sectors are considered separately,
and the eigenvalues of $\hat D$ could be found easily up to flux
line numbers of order $j=10$. For example, considering only
the ${\cal O}^{(6)}$-invariant sector, solution of the eigenvalue problem for
$j=1,2,3,\dots$ requires the diagonalization of square matrices
of size $1,2,5,8,14,20,30,40,\dots$, to be compared with a total
number of states $5,15,34,65,111,175,260,369,\dots$, if the
${\cal O}^{(6)}\times T$-action is not taken into account. 
We also found that
on these subsectors of Hilbert space, all eigenvalues already occur 
in the invariant $A_{1}^{-}$-sector, and are non-degenerate, that is, 
their corresponding eigenvectors are automatically orthogonal.
Whether the ${\cal O}$-invariant sector is also distinguished on
physical grounds depends on how the continuum limit of the lattice
theory is taken, and on how the diffeomorphism symmetry is realized,
both of which are still unresolved issues.

\vskip1.5cm
\line{\ch 4 Summary\hfil}

We have explained the need for the condition ${\det E}>0$ on physical
states in connection gravity, both classically and quantum-mechanically.
Since the spectrum of the local lattice operator $\hat {\det E}(n)$ 
is discrete, there is no problem
in principle in eliminating states with negative or vanishing
eigenvalue of $\hat {\det E}$. Eigenvalues come in pairs of opposite 
sign, and the corresponding eigenstates are related by complex conjugation.
Eigenstates of a definite sign can be constructed once the eigenstates
of $(\hat {\det E})^2$ are known. 

These considerations make the evaluation of the spectrum of $\hat {\det E}$
even more urgent, apart from its central importance 
as an ingredient in kinematical and dynamical operators
in canonical quantum gravity.
We were able to make progress in this task by taking into account 
superselection sectors related to the symmetry properties of
$\hat {\det E}(n)$ under the action of the cubic group and space reflection.

\vskip1.5cm
\line{\ch References\hfil}

\item{[1]} A. Ashtekar, Phys. Rev. Lett. 57 (1986) 2244; 
  Phys. Rev. D36 (1987) 1587.
  
\item{[2]} J.F. Barbero G., Phys. Rev. D51 (1995) 5507.

\item{[3]} R. Loll, Nucl. Phys. B460 (1996) 143.

\item{[4]} C. Rovelli and L. Smolin, Nucl. Phys. B442 (1995) 593, 
  Err. ibid. B456 (1995) 753;
  R. De Pietri and C. Rovelli, Phys. Rev. D54 (1996) 2664.

\item{[5]} A. Ashtekar and J. Lewandowski, J. Geom. Phys. 17 (1995) 191;
  J. Lewandowski, Warsaw U. preprint (1996). 

\item{[6]} T. Thiemann, Harvard U. preprint HUTMP-96/B-353 (1996).

\item{[7]} R. Loll, MPI Potsdam preprint AEI-023 (1996).
 
\item{[8]} R. Loll, Phys. Rev. D54 (1996) 5381.

\item{[9]} R. Loll, Phys. Rev. Lett. 75 (1995) 3048.

\item{[10]} T. Janssen, Crystallographic groups (North-Holland, 
  Amsterdam, 1973); S.L. Altmann, in: Quantum Theory II, ed.
  D.R. Bates (Academic Press, 1962).

\item{[11]} B. Berg and A. Billoire, Nucl. Phys. B221 (1983) 109.
  
\item{[12]} R. Loll, to be published.

\end